\documentclass[10pt,conference]{IEEEtran} 

\usepackage[utf8]{inputenc}
\usepackage[T1]{fontenc}
\usepackage[x11names,svgnames,usenames,dvipsnames,table,rgb]{xcolor}
\usepackage[english]{babel}
\usepackage{amsmath, amsfonts, amssymb}
\usepackage{csquotes}
\usepackage{graphicx}
\usepackage[binary-units=true, detect-all=true]{siunitx}
\usepackage{yquant}
\yquantset{register/default name=$\ket{\reg_{\idx}}$}
\usepackage{mathtools,amssymb,amsthm,amsfonts,nicefrac,amsmath,braket,bbm}
\usepackage{physics}
\usepackage{tikz}
\usetikzlibrary{quantikz2}
\usepackage{comment}
\usepackage{xcolor}
\usepackage{tabularx,booktabs,multirow}
\usepackage[english]{babel}
\usepackage[nottoc]{tocbibind}

\usepackage{csvsimple}
\makeatletter
\let\MYcaption\@makecaption
\makeatother
\usepackage{subcaption}
\makeatletter
\let\@makecaption\MYcaption
\makeatother

\usepackage{enumitem}

\usepackage{xargs}	
\usepackage[colorinlistoftodos,prependcaption,textsize=small]{todonotes}
\newcommandx{\unsure}[2][1=]{\todo[linecolor=red,backgroundcolor=red!25,bordercolor=red,#1]{#2}}
\newcommandx{\change}[2][1=]{\todo[linecolor=blue,backgroundcolor=blue!25,bordercolor=blue,#1]{#2}}
\newcommandx{\info}[2][1=]{\todo[linecolor=OliveGreen,backgroundcolor=OliveGreen!25,bordercolor=OliveGreen,#1]{#2}}
\newcommandx{\improvement}[2][1=]{\todo[linecolor=Plum,backgroundcolor=Plum!25,bordercolor=Plum,#1]{#2}}

\usepackage[hidelinks]{hyperref}

\hypersetup{
    unicode=false,          
    pdftoolbar=false,        
    pdfmenubar=true,        
    pdffitwindow=false,     
    pdfstartview={FitH},    
    pdftitle={},    
    pdfauthor={Viet T. Tran, Richard Kueng},     
    pdfsubject={One, Two, Three: One empirical evaluation of a two-copy shadow tomography scheme with triple efficiency},   
    pdfcreator={},   
    pdfproducer={}, 
    pdfkeywords={}, 
    pdfnewwindow=true,      
    colorlinks=true,       
    linkcolor=red,          
    citecolor=blue,        
    filecolor=teal,      
    urlcolor=blue           
}

\addto\captionsenglish{%
}
\addto\extrasenglish{%
}

\addto\captionsenglish{%
}
\addto\extrasenglish{%
}

\addto\captionsenglish{%
  \def\figurename{Fig.\!}%
}
\addto\extrasenglish{%
}

   \makeatletter
    \newcommand{\linebreakand}{%
      \end{@IEEEauthorhalign}
      \hfill\mbox{}\par
      \mbox{}\hfill\begin{@IEEEauthorhalign}
    }
    \makeatother
\makeatletter
\newcommand{\newlineauthors}{%
  \end{@IEEEauthorhalign}\hfill\mbox{}\par
  \mbox{}\hfill\begin{@IEEEauthorhalign}
  }
  
\makeatother

\makeatletter
\def\tagform@#1{\maketag@@@{\ignorespaces#1\unskip\@@italiccorr}}
\let\orgtheequation\theequation
\def\theequation{(\orgtheequation)}
\makeatother
\let\orgautoref\autoref
\renewcommand{\autoref}[1]{\def\equationautorefname{Eq.}\orgautoref{#1}}
\renewcommand{\baselinestretch}{1.0}

\usepackage{newfloat}
\usepackage{algpseudocode}
\usepackage{algorithm}
\usepackage{algorithmicx}

\algrenewcommand{\algorithmicindent}{1.0em}
\newfloat{subroutine}{htbp}{loa}
\floatname{subroutine}{Subroutine}
\makeatletter\newcommand\l@subroutine{\@dottedtocline{1}{1.5em}{2.3em}}\makeatother

\def\O{\mathcal{O}}

\DeclareMathOperator{\sgn}{sign}

\begin{document}

\title{
    One, Two, Three: One Empirical Evaluation of a Two-Copy Shadow Tomography Scheme with Triple Efficiency
}

\author{
    \IEEEauthorblockN{Viet T. Tran}
    \IEEEauthorblockA{\textit{Department of Quantum Information and Computation} \\
        \textit{at Kepler (QUICK), Johannes Kepler University Linz}\\
        Austria \\
        viet\_thuong.tran@jku.at}
    \and
    \IEEEauthorblockN{Richard Kueng}
    \IEEEauthorblockA{\textit{Department for Quantum Information and Computation} \\
        \textit{at Kepler (QUICK), Johannes Kepler University Linz}\\
        Austria \\
        richard.kueng@jku.at}

}

\maketitle

\begin{abstract}
    Shadow tomography protocols have recently emerged as powerful tools for efficient quantum state learning, aiming to reconstruct expectation values of observables with considerably fewer resources than traditional quantum state tomography. For the particular case of estimating Pauli observables, entangling two-copy measurement schemes can offer an exponential improvement in sample complexity over any single-copy strategy conceivable \cite[Huang, Kueng, Preskill, PRL (2021)]{HKP21}. A recent refinement of these ideas by King et al. \cite[King, Gosset, Kothari, Babbush, SODA (2025)]{king_triply_2025} does not only achieve polynomial sample complexity, but also maintains reasonable computational demands and utilizes joint measurements on only a small constant number of state copies. This `triple efficiency' is achievable for any subset of $n$-qubit Pauli observables, whereas single-copy strategies can only be efficient if the Pauli observables in question have advantageous structure.
    \newline
    In this work, we complement existing theoretical performance guarantees with the empirical evaluation of triply efficient shadow tomography using classical, noise-free simulations.
    Our findings indicate that the empirical sample complexity aligns closely with theoretical predictions for
    stabilizer states and, notably, demonstrates slightly improved scaling for random Gibbs states compared to established theoretical bounds. In addition, we improve a central subroutine in the triply-efficient shadow protocol by leveraging insights from a refined quantum and quantum-inspired convex optimization algorithm \cite[Henze et al. arXiv:2502.15426 (2025)]{henze_solving_2025}[Henze et al. arXiv:2502.15426 (2025)].
    \newline
    To summarize, our empirical sample complexity studies of triply efficient shadow tomography not only confirm existing theoretical scaling behavior, but also showcase that the actual constants involved are comparatively benign. Hence, this protocol has the potential to also be very sample-efficient in practice.
\end{abstract}


\section{Introduction}
\label{sec:intro}
\begin{figure*}[t]
    \centering
    \begin{subfigure}[b]{\textwidth}
        \includegraphics[width=\textwidth]{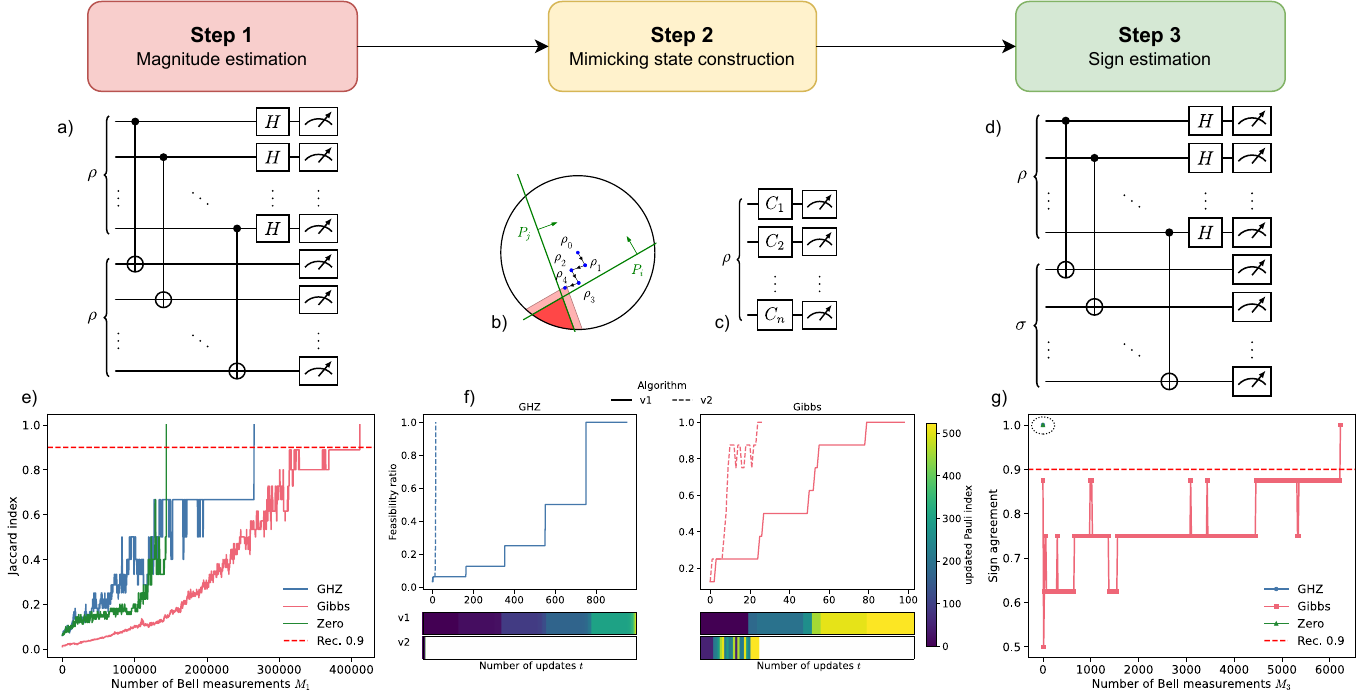}
    \end{subfigure}
    \caption{\emph{Schematic and empirical results for triply-efficient Pauli shadow tomography.}
        The protocol proceeds in three steps:
        (a) Magnitudes of Pauli obervables are estimated via repeated two-copy Bell measurements on $\rho \otimes \rho$.
        (b,c) A matrix multiplicative weights (MMW) algorithm solves a semidefinite feasibility problem to construct a mimicking state $\sigma$, guided by single-copy Pauli measurements on $\rho$.
        (d) Bell measurements on $\rho \otimes \sigma$ determine the signs of significant Pauli observables, completing the full prediction of Pauli observables (sign$\times$magnitude).
        (e) Performance of Magnitude estimation for target accuracy$\epsilon = 0.07$ for GHZ, Zero, and Gibbs states on $n=5$ qubits. The y-axis measures alignment between the support of dominant reconstructed magnitudes and the underlying ground truth. Due to its unstructured nature, the Gibbs state needs considerably more samples to surpass the 0.9 recovery threshold than the highly structured other states.
        (f) Our improved MMW update (v2) converges in 16 steps for GHZ and 33 for the Gibbs state, compared to 951 and 98 steps, respectively, for the original version (v1). The Zero state's trajectory is omitted, as it is identical to the GHZ state in terms of update count and feasibility progress, differing only in the specific Pauli indices updated.
        (g) Sign recovery for the same states as in (e). Stabilizer states (GHZ and Zero) reach perfect recovery after a single Bell measurement on $\rho \otimes \sigma$ (see dashed circle), while the Gibbs state requires significantly more samples due to its unstructured Pauli support.}
    \label{fig:three_step_process}
\end{figure*} 

Quantum state learning is the task to extract valuable information about an unknown quantum system from empirical observations. These empirical observations come in the form of quantum mechanical measurements that irreversibly disturb the quantum system in question  (`collapse of the wavefunction'). So, to obtain new data, an identical quantum system must first be prepared again from scratch and can then be measured. The total number of quantum state copies required to accurately learn the desired features is called the \emph{sample complexity} and affects all other computational resources involved (e.g. classical memory, classical and quantum runtime).

Quantum state tomography~\cite{BCG13} offers a complete solution to any quantum state learning task by reconstructing a full classical description of the density matrix $\rho$ that describes the quantum system in question. By construction, this provides us with everything there is to know about this quantum system, but at a steep cost. The density matrix $\rho$ of a $n$-qubit quantum system is a Hermitian $2^n \times 2^n$ matrix. Even the most sample-efficient state tomography approaches known to date require exponentially many state copies in $n$~\cite{KRT17,DW16,HHJWN16,FBK21}. This is not a coincidence. Fundamental lower bounds confirm this scaling both for entangling multi-copy measurements~\cite{HHJWN16} and for sequential single-copy measurements~\cite{AL25}.

Up to our knowledge it was Scott Aaronson~\cite{Aar18} who first pointed out that quantum state tomography is typically a means to an end: predict (many) features of the unknown quantum state. He then proposed to bypass the necessarily hard task of fully reconstructing $\rho$ and directly predict features, more precisely: predict many linear observables $\tr\left(O_l \rho\right)$ with $l=1,\ldots,L$. This process is now known as \emph{shadow tomography}~\cite{Aar18} or \emph{quantum data analysis}~\cite{BD21}, respectively. Broadly speaking, these protocols $\epsilon$-approximate $L$ target observables based on highly entangling quantum operations that act jointly on only $\tilde{\O} \left(n\log^2 (L)/\epsilon^4\right)$ state copies of $\rho$. Remarkably, this sample complexity is valid for any quantum state and for any observables.
The downside is that these protocols are extremely demanding in terms of quantum resources (they require exponentially deep quantum circuits in $n$ that act on a huge quantum register that hosts all state copies in parallel).

Such extreme (quantum and classical) computation demands can be circumvent if we start imposing restrictions on either the set of states involved or the collection of observables in question.
Matrix product state tomography~\cite{CPF+10}, neural network state tomography~\cite{TMC+18} and (high temperature) Gibbs state tomography~\cite{HKT22,RFOW24} are examples where one restricts the class of states. Classical shadows~\cite{MD19,PK19,HKP20} are a prominent example where one instead restricts the class of target observables.
A particularly important subclass of $n$-qubit quantum obervables are Pauli observables, i.e.\ $n$-fold tensor products of single-qubit Pauli matrices.
These are vital in quantum many-body physics, but also occur when one maps fermionic problems, which feature prominently in quantum chemistry, onto a $n$-qubit system.
The task of learning Pauli observables from empirical observations is also known as \emph{Pauli shadow tomography}. It is an interesting problem with a rich structure that is intimately related to the commutation and anticommutation relations between the Pauli observables in question.
Pauli grouping strategies~\cite{VTI20} attempt to divide the full set of target Pauli observables into a collection of ideally few subset that only contain commuting Pauli observables that can then be jointly measured. For local Pauli observables, the aforementioned classical shadows are efficient and easy to execute. In addition, de-randomized versions of classical shadows~\cite{HKP21b,GK25,KKK+24} also perform very well for structured sets of non-local Pauli observables, like the ones that arise in qubit encodings of quantum chemistry problems.
Importantly, all these methods use sequential measurements on individual copies of $\rho$, like the circuit visualized in the central panel of Fig.~\ref{fig:three_step_process} and cannot handle general collections of Pauli observables which do not feature any advantageous structure.
Again, this is not a coincidence. Fundamental lower bounds assert that any  Pauli estimation strategy that is based on sequential single-copy measurements must require exponentially many state copies in general~\cite{HKP21,CCHL21,HBC+22}.

The only way to circumvent these stringent no-go results is to change the measurement paradigm. Joint measurements on multiple state copies, like the circuit illustrated in panel a) of Fig.~\ref{fig:three_step_process}, circumvent these lower bounds and can achieve tractable sample complexities that only scale polynomially in the number of qubits $n$. Over the past years, several such strategies have been developed~\cite{HKP21,HBC+22,CGY24,king_triply_2025}. Remarkably, the most refined protocols among them achieve sample complexities comparable to general shadow tomography, but only require joint measurements on (at most) two state copies.

We will review the general idea behind such multi-copy Pauli shadow tomography approaches further below in Sec.~\ref{sec:core_concepts}. Here, we content ourselves with emphasizing that these protocols strive to achieve three desirable properties: (i) a benign scaling in sample complexity, (ii) a small number of state copies that have to be jointly measured, and, ideally,
(iii) runtime- and memory-efficient classical co-processing.
Last year, two independent research collaborations~\cite{CGY24,king_triply_2025} succeeded in achieving at least two of these three desiderata and Ref.~\cite{king_triply_2025} coined the term \emph{triply efficient shadow tomography}.
Both articles support their findings with theoretical performance guarantees, but so far none of these protocols have been actually implemented and tested empirically.
Here, we fill this gap and provide a detailed empirical analysis of the performance of triply-efficient shadow tomography based on classical simulations of both the quantum measurement stage and the classical co-processing stage.
As an added benefit, we also improve a crucial subroutine in the existing protocols by leveraging insights from a refined quantum-inspired convex optimization algorithm~\cite{henze_solving_2025} that we developed together with F.\ Henze, B.\ Ostermann, T. de Wolff and D.\ Gross.

A compact summary of our findings is illustrated in Fig.~\ref{fig:three_step_process}.
The rest of this article is organized as follows.
Sec.~\ref{sec:core_concepts} introduces core concepts and provides a conceptual foundation for understanding Pauli observable estimation and two-copy shadow tomography. Sec.~\ref{sec:methods} then elaborates on our methodological approach, including an overview of the triply efficient shadow tomography protocol, improvements to the classical co-processing stage, and details on our numerical benchmarking setup. In Sec.~\ref{sec:evaluations}, we present empirical evaluations and discuss our findings regarding the performance of each stage of the shadow tomography protocol, emphasizing comparisons between theoretical predictions and our numerical results. Additionally, we contrast the performance of the original co-processing algorithm (v1) against our improved adaptive variant (v2). Sec.~\ref{sec:conclusion} concludes this work with a succinct summary and also outlines potential avenues for future research and practical enhancements.

\section{Core concepts and technical foundation}
\label{sec:core_concepts}

\subsection{Challenge: Pauli observable estimation}

Let $S \subseteq \left\{I,X,Y,Z\right\}^{\otimes n}$ be a subset of $n$-qubit Pauli observables 
and let $\rho$ be a $n$-qubit quantum state.
The task is to approximate each Pauli observable $\tr \left( P \rho \right)$ with $P \in S$ up to additive accuracy $\epsilon$ with high probability.
Such Pauli observable estimation problems frequently arises in quantum many-body physics and quantum chemistry.

The most straightforward solution is to sequentially measure all Pauli observables $P \in S$.
This approach achieves a sample complexity of $\mathcal{O} \left( |S|/\epsilon^2 \right)$ and only requires single-copy measurements of $\rho$ in different Pauli bases -- an easy task for most existing quantum hardware platforms. 
Here, the $1/\epsilon^2$-scaling takes into account shot noise and the cardinality $|S|$ arises from sequentially measuring one Pauli observable at a time.
Substantial improvements are possible if $S$ is comprised of many Pauli observables that can be jointly measured with the same Pauli basis measurement (e.g. $Z \otimes I$, $I \otimes Z$ and $Z \otimes Z$ for $n=2$), or at least commute with each other and can therefore also be jointly measured in principle (e.g. $X \otimes Y$ and $Y \otimes X$ for $n=2$).
These joint measurements may, however, require the execution of general $n$-qubit Clifford circuits prior to measuring the individual qubits in the computational basis. And although the size of such global Clifford circuits is at most quadratic in $n$~\cite{SK14}, executing them is still very demanding on current and future hardware.
To make matters worse, commutation itself is already a rare feature among Pauli observables. To see this, note that two $n$-qubit Pauli observables $P$ and $Q$ either commute or anti-commute:
\begin{equation}
    P Q = \pm QP. \label{eq:anticommutation}
\end{equation}
And a simple counting argument reveals that a fixed $P$ admits exactly $4^n/2^{\mathrm{weight}(P)}$ distinct Pauli observables $Q$ such that $PQ=+QP$. Here, the Pauli weight $\mathrm{weight}(P) \in \left\{0,1\ldots,n\right\}$ counts the number of non-identity Pauli matrices in $P$. Unless the Pauli weight is very small, this ratio is tiny when compared to the total number of Pauli observables $4^n$.
This basic counting argument seems to limit the potential for joint measurement strategies, because quantum mechanical uncertainty relations~\cite{CBTW17}, like the famous Heisenberg uncertainty relation, tell us that anti-commuting observables are incommensurable and cannot be jointly measured.
In the following we review how a sequence of works managed to bypass this fundamental challenge by changing the underlying measurement paradigm: one has to move from sequential single-copy measurements to joint measurements on two copies of $\rho$.

\subsection{Efficient estimation of Pauli observable magnitudes}\label{subsec:bell_sampling}

Up to our knowledge, the following insight has first been presented and used in Ref.~\cite{HKP21}.
Eq.~\eqref{eq:anticommutation} highlights a fundamental feature of Pauli observables: they either perfectly commute or they perfectly anti-commute.
And this has profound implications for the two-fold Kronecker product of two general $n$-qubit Pauli observables $P$ and $Q$:
\begin{align}
    \left( P \otimes P \right) \left( Q \otimes Q \right)
    = & \left( P Q\right) \otimes \left( P Q \right) = \left( \pm Q P \right) \otimes \left( \pm Q P \right) \nonumber \\
    = & \left( \pm 1 \right)^2 (QP) \otimes (QP)                                                                       \\
    = & \left(Q \otimes Q \right) \left( P \otimes P \right).\nonumber
\end{align}
In words, all $4^n$ Kronecker products $P \otimes P$ -- which are formally Pauli observables on $2n$ qubits -- commute with each other and can therefore be jointly measured.
Now, something interesting happens if we compute the expectation value of $P \otimes P$ for two copies $\rho \otimes \rho$ of the unknown quantum state:
\begin{equation}
    \tr \left( (P \otimes P) (\rho \otimes \rho) \right) = \tr \left(P \rho \right)^2 = \left| \tr \left( P \rho \right) \right|^2.
\end{equation}
Hence, a very specific measurement on two copies of $\rho$ allows us to jointly approximate the squared moduli of \emph{all} Pauli observables!
Even more remarkably, the $2n$-qubit unitary circuit that diagonalizes all these Pauli Kronecker products is remarkably simple and structured. It corresponds to a parallel execution of $n$ two-qubit Bell basis measurements, see Fig.~\ref{fig:three_step_process} panel a) for a visual illustration. This process of executing two-copy measurements is also called \emph{Bell sampling}.
A standard concentration argument in Ref.~\cite{HKP21} showcases that a total of $\O\left( \log (|S|)/\epsilon^4\right)$ independent repetitions of this two-copy measurement circuit suffice to $\epsilon$-approximate all Pauli magnitudes $|\tr(P \rho)|$ for $P \in S$ with high probability. Here, the $1/\epsilon^4$-scaling arises from the fact that we actually approximate the square $|\tr(P \rho)|^2$ in a shot-noise limited fashion and then take a square root (which amplifies approximation errors), while the $\log (|S|)$-scaling comes from having to control the maximum error over $|S|$ different empirical magnitude approximations.

\subsection{Efficient estimation of Pauli observable signs} \label{subsec:sign_estimation}
The ability to efficiently approximate all Pauli magnitudes from a single two-copy measurement setting is a promising starting point for Pauli shadow tomography. But, approximately knowing $|\tr(P\rho)|$ cannot yet provide information about the actual sign ($\pm1$) associated with this particular Pauli expectation value.
Ref.~\cite{HKP21} proposed to recover these signs based on sequential thresholding and direct single-copy Pauli measurements.
More precisely, we first use two copy measurements to jointly approximate $|\tr(P \rho)|$ for all $P \in S$. Then, we sort out all small Pauli expectation values: whenever our estimate $\hat{u}_P$ for the absolute value $u_P=|\tr(P \rho)|$ obeys $\hat{u}_p < 3\epsilon/4$, we simply set this observable to $0$ (thresholding). This leaves us with an estimated set $\hat{S}_\epsilon \subseteq S$ of significant Pauli observables, where each element $P \in S_\epsilon$ obeys $\hat{u}_P \geq 3\epsilon/4$. And for these observables, we can sequent to obtain the missing sign information (and, potentially, a more accurate statistical approximation).
As already discussed above, the cost of this sequential Pauli observable estimation strategy scales as $\mathcal{O} \left( |S_\epsilon| \log (|S_\epsilon|/\delta)/\epsilon^2\right)$, where $S_\epsilon$ is the true significant Pauli support of $\rho$. This sample complexity remains favorable if the number of significant Pauli expectation values remains bounded.
Typical quantum states have this property, but for large sets $S$ and certain structured quantum states, $|S_\epsilon|$ can become exponentially large in $n$. A concrete and illustrative example for this behavior are stabilizer states, such as computational basis states or the GHZ state. By definition, a $n$-qubit stabilizer state $\rho = |\psi \rangle \! \langle \psi|$ obeys $| \tr(P \rho)|=1$ for exactly $2^n$ Pauli observables and $\tr(P \rho)=0$ otherwise. Consequently, $|S_\epsilon|=2^n$ for every $\epsilon \in [0,1)$.

Subsequent work by Chen, Gong and Ye on the one hand~\cite{CGY24}, and King, Gosset, Kothari and Babbush on the other hand~\cite{king_triply_2025}, found a way to circumvent this issue of having to sequentially measuring all significant Pauli observables to recover the missing sign information. Here, we follow the second approach which the authors termed \emph{triply efficient (Pauli) shadow tomography}. The key idea is to include an additional data co-processing stage, where one solves the following feasibility problem over the convex set of $n$-qubit quantum states, i.e.\ the set of complex-valued $2^n \times 2^n$ matrices that are Hermitian ($\sigma^\dagger =\sigma$), positive semidefinite ($\sigma \succeq 0$) and have unit trace ($\tr(\sigma)=1$):
\begin{align}
    \begin{split}
        \text{find}       & \quad \sigma \in \mathbb{C}^{2^n \times 2^n}  \label{eq:feasibility-problem}                         \\
        \text{subject to} & \quad \left|\tr\left(P \sigma \right)\right| \geq \epsilon/4 \quad \text{for all $P \in S_\epsilon$} \\
                          & \quad \sigma^\dagger = \sigma, \; \sigma \succeq 0, \; \tr(\sigma)=1.
    \end{split}
\end{align}
In words: we are looking for a valid $n$-qubit density matrix $\sigma$ that achieves large Pauli observable magnitudes at precisely the same locations as the unknown density matrix $\rho$. The authors of Ref.~\cite{king_triply_2025} call such a state $\sigma$ a \emph{mimicking state}.
Such a $\sigma$ must always exist, because the unknown underlying density matrix $\rho$ would solve Problem~\eqref{eq:feasibility-problem} by construction.
But insisting to find $\rho$ here would be akin to full quantum state tomography and too resource-intensive. Fortunately, this is also not necessary.
Instead, we use any density matrix $\sigma$ that solves Problem~\eqref{eq:feasibility-problem} and determine a way to prepare the associated $n$-qubit quantum state in a quantum computer\footnote{Note that the task of preparing the mimicking state $\sigma$ on a quantum computer is daunting and can be expensive, both in quantum circuit size and the classical cost associated with transpilation. The particular algorithm we propose here for finding $\sigma$ is very beneficial in this regard, because it updates quantum Gibbs states. And there are explicit, if abstract, quantum circuit constructions to prepare such Gibbs states, see e.g.~\cite{TOVPV11,RFA24}.}. This together with access to individual copies of $\rho$ allows us to prepare independent copies of the joint $2n$-qubit state $\rho \otimes \sigma$ on our quantum computer. And subsequent Bell-basis measurements then allow us to jointly approximate \emph{all} Pauli tensor product observables $P \otimes P$ evaluated on $\rho \otimes \sigma$. And here comes the interesting part:  each such expectation value can be simplified to
\begin{equation}
    \tr \left( (P \otimes P) (\rho \otimes \sigma \right)
    = \sgn\left( \tr(P \rho) \right) \left| \tr(P \rho) \right| \tr(P\sigma).
\end{equation}
And since we already have approximated $|\tr(P \rho)|$ earlier and we know $\tr(P \sigma)$ exactly from the feasibility problem, we can now approximate the missing sign information:
\begin{equation}
    \sgn\left( \tr \left( P\rho \right) \right\} =\frac{\tr \left( (P \otimes P) \rho \otimes \sigma \right)}{\left| \tr(P \rho) \right| \tr(P \sigma)} \quad \text{for all $P \in S_\epsilon$}. \label{eq:sign_estimation}
\end{equation}
Joint Bell-basis type measurements together with the constraint that both $|\tr(P \rho)|$ and $|\tr(P \sigma)$ have at least magnitude $\epsilon/4$ now ensure that a total of $\mathcal{O} \left( \log \left( |S_\epsilon|\right)/\epsilon^4 \right)$ Bell basis measurement repetitions provide enough statistical data to correctly determine each $\sgn(P\rho)$. The $1/\epsilon^4$-scaling now comes from the fact that $\tr( (P \otimes P) \rho \otimes \sigma)$ has (at least) the same magnitude as $\epsilon^2 \sgn\left\{ \tr(P \rho)\right\}$ and an accuracy of order $\epsilon^2$ is required to resolve the underlying signs. Shot noise then amplifies this quadratic accuracy demand to order $1/\epsilon^4$. Finally, the $\log (|S_\epsilon|)$-scaling comes from having to control the maximum deviation error over $|S_\epsilon|$ different sign approximations.
Needless to say, the set of significant Pauli observables $S_\epsilon$ can never be larger than the full set of target Pauli observables $S$ and is often much smaller. However, $|S_\epsilon| \leq |S|$ already ensures that the cost of this paralellized sign estimation is at worst comparable to the cost of the first stage (magnitude estimation). So, the overall sample complexity scaling is $\mathcal{O} \left( \log \left( |S| \right)/\epsilon^4\right)$ which can be exponentially better in $|S_\epsilon|$ than the sample-complexity $\mathcal{O} \left( |S_\epsilon|/\epsilon^4 \right)$ achieved by sequentially measuring all significant Pauli observables directly to determine the signs.

\section{Methods}\label{sec:methods}

In Sub.~\ref{subsec:triply_eff}, we begin by recapitulating the three central steps of the triply efficient shadow tomography protocol proposed by \cite{king_triply_2025} . In Sub.~\ref{subsec:mm_state_hu_algo}, we then focus on the second step (mimicking-state construction) and show how this crucial subroutine can be accelerated using recent advances in quantum-inspired convex optimization.
Finally in Sub.~\ref{subsec:numerical_details}, we describe our numerical implementation, detailing the class of quantum states under study and the performance metrics used to evaluate the results.
\subsection{Overview: three-step protocol}\label{subsec:triply_eff}
As already explained in Sec.~\ref{sec:core_concepts}, the triply efficient shadow tomography protocol \cite{king_triply_2025} further improves earlier two-step Bell sampling measurement protocols by introducing an additional intermediate step to construct a mimicking state $\sigma$.

We now show that access to such a mimicking state allows for efficiently estimating \emph{every} $\tr(P \rho)$ conceivable with  quasi-linear sample complexity in qubit size $n$.

The three stages are:
\begin{enumerate}[leftmargin=*, label=\textbf{\arabic*.}]
    \item \textbf{Stage -- Magnitude estimation:} Repeat Bell basis measurements (see Sub.~\ref{subsec:bell_sampling}) on two state copies $\rho \otimes \rho$ to obtain approximations of $u_P=|\tr(P \rho)|$ for each $P \in S$ with sample complexity $O(\log|S| / \epsilon^4)$ and accuracy $\epsilon/4$.

    \item \textbf{Stage -- Mimicking state construction:} For a desired approximation accuracy $\epsilon$, define the (estimated) support of large Pauli observables
          \begin{align}
              S_\epsilon := \left\{P \in S : u_p=| \mathrm{tr}(P \rho)| \geq 3\epsilon/4 \right\},
          \end{align}
          based on the (estimated) $u_P$s from Stage~1.
          Then construct a \emph{mimicking state} $\sigma$ such that all Paulis $P$ in the support $S_\epsilon$ of large Pauli observables also have a significant magnitude $|\tr(P \sigma)| \geq \epsilon/4$. This step can be formalized as a  feasibility problem, see Problem~\ref{eq:feasibility-problem}. The solution algorithm, which we discuss in more detail in Sub.~\ref{subsec:mm_state_hu_algo} below, performs $\mathcal{O}(n/\epsilon^2)$ steps and requires  $\mathcal{O}(\log(n/\epsilon)/\epsilon^2)$ single-copy measurements of $\rho$ in each step.

    \item \textbf{Stage -- Sign estimation:} Perform Bell basis measurements (see Sub.~\ref{subsec:bell_sampling}) on the joint state $\rho \otimes \sigma$ to estimate $\tr(P \rho) \tr(P \sigma)$ for each $P \in S_\epsilon$. Since we know $\sigma$, and can by extension classically compute $\tr(P \sigma)$, these measurements yield the sign of $\tr(P \rho)$. This stage takes $\mathcal{O}(\log|S|/\epsilon^4)$ Bell samples of the composite system $\rho \otimes \sigma$.
\end{enumerate}
\noindent
Each stage is sample-efficient and either uses single-copy Pauli measurements (Stage 2) or two-copy Bell basis measurements that can be decomposed into a parallel layer of $n$ two-qubit CNOT gates and an additional layer of $n$ single-qubit Hadamard gates. This avoids the need for high-degree entangled measurements. The protocol achieves total sample complexity of order
\begin{align}
    \O\left( n\log(n/\epsilon)/\epsilon^4 \right)
\end{align}

Having outlined the three primary stages of the triply efficient shadow tomography protocol, we now turn our attention to the key computational bottleneck: find a mimicking state $\sigma$ in Stage 2.
We execute this feasibility search by leveraging the matrix multiplicative weights (MMW) algorithm~\cite{arora_multiplicative_2012} and enhance it further with a quantum-inspired variation called \emph{Hamiltonian updates} \cite{brandao_faster_2022,henze_solving_2025}.

\subsection{Mimicking state construction via Hamiltonian updates}\label{subsec:mm_state_hu_algo}
The overarching goal of the mimicking state construction can be stated as a feasibility problem over the convex set of $n$-qubit density matrices. We are given estimates $u_P$ of the magnitude $|\tr(P\rho)|$ which satisfy $|u_P - |\tr(P\rho)|| \leq \epsilon/4$ with (very) high probability. From these magnitudes, we determine the support $S_\epsilon$ of significant Pauli observables.
We want to find a mimicking state $\sigma$ that mimics $\rho$ such that it has the same significant Pauli observables with a threshold of $\epsilon/4$. Formally, this is stated as a feasibility problem~\ref{eq:feasibility-problem}, that features a large collection of non-convex constraints $|\tr(P \sigma)| \geq \epsilon/4$.
If all these constraints are met, we know that we have found a correct mimicking state $\sigma$.
Else if this is not the case, we must be able to find a $P \in S$ such that $| \tr (P \sigma) | < \epsilon/4$. Subsequently, we can linearize this one constraint by first performing Pauli basis measurements on $\rho$ to approximate the corresponding sign $r_P = \mathrm{sign} \left( \tr (P \rho) \right)$ of the true state $\rho$ and using this `correct sign' to replace $| \tr (P \sigma)| \geq \epsilon/4$ by $r_P \tr (P \sigma) \geq \epsilon/4$.
This linearized re-formulation of a violated feasibility constraint can then be used to update a currently infeasible candidate mimicking state $\sigma$ to make it less infeasible. Iterating this basic idea gives rise to a matrix multiplicative weight (MMW) type algorithm ~\cite{arora_multiplicative_2012}.
This is a classical optimization algorithm that, in quantum-inspired terminology, iteratively refines the candidate solution $\sigma$ by updating the Hamiltonian matrix $H$ associated to a Gibbs state parametrization $\sigma = \exp \left( - \beta H \right)/\tr\left( \exp \left( - \beta H \right) \right)$. The iterative procedure in Ref.~\cite{king_triply_2025} starts with $H=0$ and, consequently, $\sigma = \mathbb{I}/2^n$. It then checks whether both $|\tr(P\sigma) - u_P| > \epsilon/2$ \textit{and} $|\tr(P\sigma) + u_P| > \epsilon/2$ hold for every $P \in S_\epsilon$.
If we don't find any such $P$, we can conclude that the current iterate $\sigma$ obeys $|\tr(P \sigma)| \geq u_p - \epsilon/2 \geq \epsilon/4$ for all $P \in S_\epsilon$, because $u_P \geq 3 \epsilon/4$ by construction. In other words: we have found a mimicking state $\sigma$. Else if we do find such a $P$, we update the Hamiltonian $H$ by adding a term $\sgn(\tr(P\sigma) - r_P u_P) P$, compute the new Gibbs state iterate and check the above conditions again. This single update rule is then repeatedly executed until we either converge to a mimicking state (if we can't find a $P$ that obeys both conditions above) or a maximum number of $T=\lceil 64n/\epsilon^2 \rceil$ iterations has been exhausted (pre-specified stopping time).
This procedure is summarized in Algorithm~\ref{alg:mm_algo_general}, where Subroutine~\ref{alg:update_step_v1} is performed for each update.

Since its inception in 2012, the MMW meta-algorithm framework has seen several adaptations and improvements. One of them is Ref.~\cite{henze_solving_2025}, where we introduce an adaptive step size $\eta$ with backtracking and choose the correction factor of $P$ proportionally to $(\tr(P\sigma) - r_P u_P)$ instead of the sign of this violation. Our modified mimicking state construction algorithm maintains the overall structure of MMW, but replaces the original update rule (Subroutine~\ref{alg:update_step_v1}) from Ref.~\cite{king_triply_2025} with a more sophisticated Subroutine~\ref{alg:update_step_v2}, called adaptive Hamiltonian Updates. We will see below that this leads to better performances across the board and can even flag inconsistency issues with the underlying support of significant Pauli observables $|S_\epsilon|$.

\begin{algorithm}[ht]
    \floatname{algorithm}{General Algorithm}
    \begin{algorithmic}[0]
        \State \textbf{Input:} A precision parameter $\epsilon\in(0,1)$, $\mathcal{O}(n\,\log(n/\epsilon)/\epsilon^4)$ copies of an unknown $n$-qubit state $\rho$, and estimates $\{u_P\}_{P\in\mathcal{P}(n)}$ s.t.
        \[
            u_P\ge 0 \quad\text{and}\quad |u_P-|\tr(P\rho)||\leq \epsilon/4 \quad \forall P\in\mathcal{P}^{(n)}
        \]
        \State \textbf{Output:} A density matrix $\sigma$ satisfying, with high probability,
        \[
            |\tr(P\sigma)|\geq \epsilon/4 \quad \forall P\in\mathcal{P}(n)\text{ with } u_P\geq 3\epsilon/4
        \]
        \State \textbf{Initialization:} Set
        \[
            \begin{split}
                T      & \gets \lceil 64n/\epsilon^2\rceil+1,\quad \beta\gets \sqrt{n/T},\quad \eta \text{ (initial step size)} \\
                \sigma & \gets \tfrac{1}{2^n}\mathbbm{1}_{2^n\times2^n},\quad H\gets 0_{2^n\times2^n}
            \end{split}
        \]
        \For{$t=0$ \textbf{to} $T-1$}
        \State \textbf{Search:} Find a Pauli $P\in\mathcal{P}(n)$ with $u_P\geq 3\epsilon/4$ s.t. both
        \[
            |\tr(P\sigma) \pm u_P|>\epsilon/2
        \]
        \If{such a Pauli $P$ is found}
        \State Use $O(\log T/\epsilon^2)$ copies of $\rho$ to estimate
        \[
            r_P=\sgn(\tr(P\rho))
        \]
        \State Perform (adaptive) update step
        \[
            H,\sigma,\eta \gets \Call{UpdateStep$_{1,2}$}{\beta,H,\sigma,P,r_P,u_P,\eta}
        \]
        \Else
        \State \Return \textbf{true}, $\sigma$ \Comment{Feasible solution found}
        \EndIf
        \EndFor
        \State \Return \textbf{false}, $\sigma$ \Comment{No feasible solution found}
    \end{algorithmic}
    \caption{Compute a mimicking state}
    \label{alg:mm_algo_general}
\end{algorithm}

In summary, our refined matrix multiplicative weights procedure, coupled with adaptive Hamiltonian updates, provides a practical way to construct the mimicking state $\sigma$ with fewer iterations in typical scenarios. Having established the theoretical framework for Stage 2, we now describe how all three stages of the protocol are implemented and evaluated in thorough numerical experiments. Specifically, we will detail the types of quantum states we test, how we measure success, and the computational considerations for each algorithmic component.

\subsection{Benchmarking setup}\label{subsec:numerical_details}

Having reviewed the triply efficient Pauli shadow tomography procedure with an improved search for mimicking states, we are now ready to outline the details of our numerical experiments.
The results of our studies can then be found in Sec.~\ref{sec:evaluations} below.

\begin{subroutine}[ht]
    \caption{UpdateStep$_1$, original update from \cite{king_triply_2025}}
    \label{alg:update_step_v1}
    \begin{algorithmic}[1]
        \State \textbf{Input:} $\beta$, current Hamiltonian $H$, current state $\sigma$, violating Pauli $P$, sign $r_P$, magnitudes value $u_P$, current step size $\eta$
        \State Set $\eta \gets 1$, \quad $\Delta H \gets \sgn(\tr(P\,\sigma) - r_P\,u_P) P$
        \State $H_{\text{new}} \gets H + \eta \Delta H$
        \State $\sigma_{\text{new}} \gets \exp(-\beta\,H_{\text{new}})/\tr(\exp(-\beta\,H_{\text{new}}))$
        \State \Return $H_{\text{new}},\,\sigma_{\text{new}},\, \eta$
    \end{algorithmic}
\end{subroutine}

\begin{subroutine}[h]
    \caption{UpdateStep$_2$, improved Hamiltonian update}
    \label{alg:update_step_v2}
    \begin{algorithmic}[1]
        \State \textbf{Input:} $\beta$, current Hamiltonian $H$, current state $\sigma$, violating Pauli $P$, sign $r_P$, magnitudes value $u_P$, current step size $\eta$
        \State Set $\delta \gets \tr(P\,\sigma) - r_P\,u_P$\,, \quad $\Delta H \gets \delta \cdot P$
        \While{true}
        \State $H_{\text{new}} \gets H + \eta \Delta H$
        \State $\sigma_{\text{new}} \gets \exp(-\beta\,H_{\text{new}})/\tr(\exp(-\beta\,H_{\text{new}}))$
        \If{$|\tr(P\sigma_{\text{new}})-r_P\,u_P| < |\delta|$}
        \State \Return $H_{\text{new}},\,\sigma_{\text{new}},\, \eta \times 1.3$ \Comment{Valid update step}
        \Else
        \State Set $\eta \gets \eta/2$ \Comment{Additional substep}
        \If{$\eta < 10^{-20}$}
        \State \Return $H,\,\sigma,\,\eta$ \Comment{Numerically insignificant}
        \EndIf
        \EndIf
        \EndWhile
    \end{algorithmic}
\end{subroutine}

\paragraph*{\textbf{0. Stage -- Generation of test states}}

We generated a diverse set of synthetic quantum states to benchmark reconstruction performance for system sizes ranging from $n=2$ to $n=7$ qubits. This included structured states such as GHZ states~\cite{greenberger_going_1989}, $( \ket{0}^{\otimes n} + \ket{1}^{\otimes n})/\sqrt{2}$, a computational basis state $\ket{0}^{\otimes n}$, as well as randomized Gibbs states of Pauli Hamiltonians. The former are stabilizer states and serve as extremal test cases, where the significant Pauli support size grows exponentially with $n$: $|S_\epsilon| = 2^n$ for any $\epsilon \in [0,1)$. In contrast, our Gibbs states capture the behavior of generic thermal states. The fact that the Hamiltonian is itself a sum of Pauli terms also allows us to interpolate the support size of significant Pauli observables (to some extent).
More precisely, we constructed the Hamiltonian as $H= \sum_{i=1}^k P_i$, where each $P_i$ was sampled uniformly from the set of all $4^n$ Pauli observables (but without replacement).
The number $k$ determines the number of Pauli terms that feature in the Hamiltonian. By construction, this also affects the number of Pauli observables with significant modulus.
To obtain a roughly equal coverage of different significant support sizes $|S_\epsilon|$ in log scale, we cover the range from $|S_\epsilon|=2^0$ to $|S_\epsilon|=4^n$ with $m=100$ specific points chosen from a logarithmic grid. More precisely,
for fixed qubit number $n$, we use the following $m=100$ values for the number of Pauli terms:
\begin{align}
    k \in & \left\{\lfloor \left(k_{\max} (n)\right)^{j/m} \rfloor: j=1,\ldots,m \right\},
\end{align}
where $k_{\max}(n)=2^{(2n-1)}$ for $n=2,3,4$ and $k_{\max}(n)=2^{(n+3)}$ for $n=5,6,7$ have been determined empirically to ensure that the generated states explore a wide range of significant Pauli support sizes $|S_\epsilon|$ while still permitting a numerically stable reconstruction all the way down to approximation accuracy $\epsilon=0.01$.
Finally, we normalized each Hamiltonian $H$ by its spectral norm $\|H \|$ to ensure a comparable size across all test instances and generated the associated test Gibbs state at unit inverse temperature $\beta =1$:
\begin{equation}
    \rho = \exp (-H/\|H\|) / \operatorname{Tr}(\exp (-H/\|H\|))
\end{equation}
All simulations were performed using Python with NumPy~\cite{harris2020array}, SciPy~\cite{2020SciPy-NMeth}, Numba~\cite{lam2015numba} and CuPy~\cite{cupy_learningsys2017} in a noise-free classical simulation.

\paragraph*{\textbf{1. Stage -- Identifying significant Pauli observables}}

The overall goal of Pauli shadow tomography is to $\epsilon$-approximate many Pauli observables $\tr (P \rho)$, where each $P$ is chosen from a (potentially very large) set $S$ of $n$-qubit Pauli observables. Recall that triply efficient shadow tomography works in three stages and Stage 1 is devoted to approximately identify the subset of significant Pauli observables.
This is achieved by repeatedly performing Bell basis measurements on independent copies of $\rho \otimes \rho$ and using the measurement statistics to jointly approximate each $u_P$ by an empirical average $\hat{u}_P$. This step is standard, but somewhat technical. We refer to Sec.~\ref{sec:core_concepts} for the overall idea and to \cite[Supplemental Material]{HKP21} for a detailed explanation of this process. What matters here is that access to all $\hat{u}_P$s for $P \in S$ allows us to approximately predict the support set of significant Pauli observables. More precisely, for $\mu \in (0,1)$, we set
\begin{equation}
    \hat{S}_\mu = \left\{ P \in S:\; \hat{u}_P \geq \mu \right\} \subseteq S.
\end{equation}
This is a data-driven prediction of the true support set $S_\mu = \left\{ P \in S:\; u_P \geq \mu \right\}$ and both sets coincide in the limit of asymptotically many Bell basis samples. But in the more realistic case of finite sample statistics, the two sets can deviate from each other. We use the \emph{Jaccard index},
\begin{align}
    J(\hat{S}_\mu, S_\mu) = \left(|\hat{S}_\mu \cap S_\mu| \right) / \left(|\hat{S}_\mu \cup S_\mu|\right),
\end{align}
to quantify the relative overlap between the two sets and asses the quality of our support prediction. The choice of the Jaccard index is motivated by its robustness and interpretable link to the $\ell_1$-distance between predicted and true supports.

In our empirical studies, we execute our simulations of Bell basis sampling simulation in measurement blocks. Since this is a simulation, we know $\rho$ and, by extension, can also compute the true set $S_{\mu}$.
This, in turn, allows us to also track the Jaccard index between the predicted support set $S_{\mathrm{pred}}$ and the true support set $S_\mu$ as the number of measurement blocks increases.
We terminate the simulation as soon as this Jaccard index exceeds a pre-defined threshold of $0.9$ for successful approximate recovery. We record the number of measurements $M$ required to achieve this threshold.
For every test state from above and re-scaled accuracy $\mu = (3/4) \epsilon \in \{0.05,0.07,0.11,0.16,0.23,0.34,0.5\}$, we repeat this adaptive support recovery process across 10 random seeds, to obtain robust statistics on the sample complexity. We also limit the maximum number of (simulated) measurements to $M_{\max,1} = 3 \times 10^{7}$ to ensure that the procedure terminates in a reasonable time frame.

\ifdefined\standalone
\documentclass{article}
\usepackage{tabularx,booktabs,multirow,amssymb, amsmath}
\usepackage[table]{xcolor}
\usepackage[binary-units=true, detect-all=true]{siunitx}
\usepackage[margin=2cm]{geometry}
\begin{document}
\fi

\definecolor{lightgray}{gray}{0.9}

\begin{table*}[htp]
    \centering

    \caption{Estimated scaling exponent $\alpha_i$ in $X = \mathcal{O}(\epsilon^{-\alpha_i})$ from log-log linear fits. Columns show estimates and corresponding percentiles.}
    \label{tab:slope_scaling}
    \begin{tabularx}{0.9\linewidth}{l X
            c >{\color{gray}}c
            c >{\color{gray}}c
            c >{\color{gray}}c
            c >{\color{gray}}c
        }
        \toprule
                                                &               & \multicolumn{2}{c}{Step 1}                                  & \multicolumn{4}{c}{Step 2}                                                                      & \multicolumn{2}{c}{Step 3}                                                                                                                              \\
                                                &               & \multicolumn{2}{c}{}                                        & \multicolumn{2}{c}{v1}                                                                          & \multicolumn{2}{c}{v2}     & \multicolumn{2}{c}{}                                                                                                       \\
        \cmidrule(lr){3-4} \cmidrule(lr){5-6} \cmidrule(lr){7-8}  \cmidrule(lr){9-10}
        qubit size $n$                          & State         & $\alpha_1$                                                  & P[2.5\,, 97.5]                                                                                  & $\alpha_2$                 & P[2.5\,, 97.5]                                              & $\alpha_3$    & P[2.5\,, 97.5] & $\alpha_4$ & P[2.5\,, 97.5] \\
        \midrule
        \multirow[c]{3}{*}{2}                   & GHZ           & 2.61                                                        & [1.75, 3.49]                                                                                    & 1.33                       & [1.33, 1.34]                                                & \textbf{0.42} & [0.41, 0.43]   & -0.09      & [-0.13, -0.05] \\
                                                & Gibbs         & 3.12                                                        & [3.06, 3.18]                                                                                    & 1.51                       & [1.49, 1.52]                                                & \textbf{0.79} & [0.78, 0.80]   & 0.82       & [0.80, 0.84]   \\
                                                & Zero          & 3.04                                                        & [1.99, 3.94]                                                                                    & 1.33                       & [1.33, 1.34]                                                & \textbf{0.42} & [0.41, 0.43]   & -0.06      & [-0.09, -0.03] \\
        \cmidrule{1-10}
        \multirow[c]{3}{*}{3}                   & GHZ           & 4.07                                                        & [3.87, 4.25]                                                                                    & 1.32                       & [1.31, 1.32]                                                & \textbf{0.43} & [0.43, 0.44]   & -0.05      & [-0.07, -0.03] \\
                                                & Gibbs         & 3.88                                                        & [3.86, 3.90]                                                                                    & 1.54                       & [1.52, 1.55]                                                & \textbf{1.07} & [1.06, 1.08]   & 1.83       & [1.79, 1.86]   \\
                                                & Zero          & 3.91                                                        & [3.74, 4.09]                                                                                    & 1.32                       & [1.31, 1.32]                                                & \textbf{0.43} & [0.43, 0.44]   & -0.03      & [-0.06, -0.01] \\
        \cmidrule{1-10}
        \multirow[c]{3}{*}{4}                   & GHZ           & 3.96                                                        & [3.82, 4.12]                                                                                    & 1.30                       & [1.30, 1.31]                                                & \textbf{0.49} & [0.48, 0.50]   & -0.10      & [-0.15, -0.07] \\
                                                & Gibbs         & 4.06                                                        & [4.04, 4.08]                                                                                    & 1.61                       & [1.59, 1.63]                                                & \textbf{1.35} & [1.34, 1.36]   & 2.24       & [2.20, 2.28]   \\
                                                & Zero          & 4.02                                                        & [3.86, 4.17]                                                                                    & 1.31                       & [1.30, 1.31]                                                & \textbf{0.49} & [0.48, 0.51]   & -0.11      & [-0.15, -0.08] \\
        \cmidrule{1-10}
        \multirow[c]{3}{*}{5}                   & GHZ           & 4.06                                                        & [3.97, 4.16]                                                                                    & 1.29                       & [1.29, 1.30]                                                & \textbf{0.51} & [0.51, 0.52]   & -0.09      & [-0.12, -0.05] \\
                                                & Gibbs         & 4.05                                                        & [4.04, 4.07]                                                                                    & 1.61                       & [1.59, 1.62]                                                & \textbf{1.52} & [1.51, 1.54]   & 2.24       & [2.20, 2.29]   \\
                                                & Zero          & 3.97                                                        & [3.83, 4.07]                                                                                    & 1.29                       & [1.29, 1.30]                                                & \textbf{0.51} & [0.51, 0.52]   & -0.10      & [-0.13, -0.07] \\
        \cmidrule{1-10}
        \multirow[c]{3}{*}{6}                   & GHZ           & 4.01                                                        & [3.92, 4.09]                                                                                    & 1.29                       & [1.28, 1.29]                                                & \textbf{0.50} & [0.49, 0.51]   & -0.04      & [-0.06, -0.02] \\
                                                & Gibbs         & 4.04                                                        & [4.03, 4.05]                                                                                    & 1.82                       & [1.79, 1.84]                                                & \textbf{1.63} & [1.62, 1.65]   & 2.33       & [2.30, 2.37]   \\
                                                & Zero          & 3.98                                                        & [3.88, 4.07]                                                                                    & 1.29                       & [1.28, 1.29]                                                & \textbf{0.50} & [0.49, 0.51]   & -0.06      & [-0.09, -0.04] \\
        \cmidrule{1-10}
        \multirow[c]{3}{*}{7}                   & GHZ           & 3.96                                                        & [3.89, 4.02]                                                                                    & 1.28                       & [1.28, 1.29]                                                & \textbf{0.39} & [0.38, 0.40]   & -0.03      & [-0.04, -0.01] \\
                                                & Gibbs         & 4.04                                                        & [4.03, 4.05]                                                                                    & 1.82                       & [1.79, 1.85]                                                & \textbf{1.29} & [1.26, 1.31]   & 2.28       & [2.24, 2.34]   \\
                                                & Zero          & 4.01                                                        & [3.96, 4.08]                                                                                    & 1.28                       & [1.28, 1.29]                                                & \textbf{0.39} & [0.37, 0.40]   & -0.02      & [-0.04, -0.01] \\
        \cmidrule{1-10}
        \multicolumn{2}{l}{Theoretical scaling} & $\alpha_1 =4$ & \textcolor{gray}{$\because\mathcal{O}(\log|S|/\epsilon^4$)} & \multicolumn{4}{c}{$\alpha_{2,3}=2$\quad \textcolor{gray}{$\because\mathcal{O}(n/\epsilon^2)$}} & $\alpha_4=4$               & \textcolor{gray}{$\because\mathcal{O}(\log|S|/\epsilon^4)$}                                                                \\
        \cmidrule{1-10}
        \bottomrule
    \end{tabularx}
\end{table*}

\ifdefined\standalone
\end{document}
\fi

\paragraph*{\textbf{2. Stage -- Mimicking state construction}}
We implement this stage using both the original algorithm from Ref.~\cite{king_triply_2025} (see Algorithm~\ref{alg:mm_algo_general} with Subroutine~\ref{alg:update_step_v1}) and our improved variant (see Algorithm~\ref{alg:mm_algo_general} with with Subroutine~\ref{alg:update_step_v2}). The magnitude estimates from the previous approximate support recovery step were used as input to both algorithms.
We track the number of iterations required for convergence to a correct mimicking state.
Again, this is possible, because we do classical simulations and therefore know $\rho$ and, by extension, $S_\epsilon$ for each test run.
For our improved algorithm, we also record the number of step size adaptations, each of which incurs an additional Gibbs state computation -- the most expensive step in the classical postprocessing.
To isolate the cost of iterations only, we replace actual sign estimation with an oracle. Instead of simulating a Pauli basis measurement of $\rho$ during each update, we use a function that returns the true sign of $\tr(P\rho)$, computed numerically.
Finally, we set the initial step size to $\eta = \tfrac{3}{8} 2^n$ in our subroutine (see Algorithm~\ref{alg:update_step_v2}). We found empirically that this particular scaling works well across all studies system sizes $n$.

\paragraph*{\textbf{3. Stage -- Pauli sign estimation}}
Given the mimicking state $\sigma$ and the previously estimated magnitudes $u_P$, we perform the final sign estimation step using Bell sampling on $\rho \otimes \sigma$ (see Sub.~\ref{subsec:bell_sampling}). This yields an estimate of the product $\tr(P \rho)\tr(P \sigma)$ for each $P \in S$.

Since we computed a classical representation of $\sigma$ in Stage 2, we can now compute each $\tr(P \sigma)$ directly.
As described in Sec.~\ref{sec:core_concepts}, this allows us now to infer the sign $r_P = \mathrm{sign} \left(\tr(P \rho)\right)$ for each $P \in \hat{S}_\epsilon$ using Eq.~\eqref{eq:sign_estimation} and a stream of Bell basis measurements on independent copies of $\rho \otimes \sigma$. The resulting empirical predictions $\hat{r}_P$ are guaranteed to converge to the true signs $r_P$ in the limit of asymptotically many samples. But errors can occur in the practically relevant regime of few samples. We quantify these errors by an appropriately normalized Mean Squared Error (MSE) over \emph{all} Pauli observables\footnote{Note that this loss function is actually the (squared) Euclidean distance between true and approximated Pauli vector of the state $\rho$, or, equivalently, the squared Hilbert-Schmidt norm difference between true quantum state $\rho=(1/2^n)\sum_P r_P P$ and its reconstructed Pauli expansion $\hat{\rho}= (1/2^n)\sum_P \hat{r}_p \hat{u}_P P$.}:
\begin{equation}
    \mathrm{MSE} = \frac{1}{2^n} \sum_{P \text{ $n$-qubit Pauli}} \left| \hat{r}_P \hat{u}_P - \tr \left( P \rho \right) \right|^2.
\end{equation}
In addition, we also track the number of Bell samples required to achieve a sign agreement of at least 90\% with the true Pauli vector. Formally, the sign agreement is defined as the ratio
\begin{equation}
    \Big(\sum_{i : |p_i| \ge \mu} \mathbbm{1}[\hat{r}_i = \sgn(p_i)] \Big)/ \Big( \sum_{i: |p_i| \ge \mu} 1 \Big) \in \left[0,1\right]
\end{equation}
and limit the number of Bell samples in this stage to $M_{\max,3} = 7 \times 10^5$ to ensure practical runtime. This upper limit is significantly smaller than the number of Bell samples typically required during the magnitude estimation stage.

The full implementation of all algorithmic components, including Gibbs state sampling, identification of significant Pauli obervables with Bell sampling, adaptive mimicking state construction, and sign reconstruction, will be made available online \cite{tran_one_2025}.

\section{Evaluations}\label{sec:evaluations}

\begin{figure}[h]
    \centering
    \begin{subfigure}[b]{0.50\textwidth}
        \includegraphics[width=\textwidth]{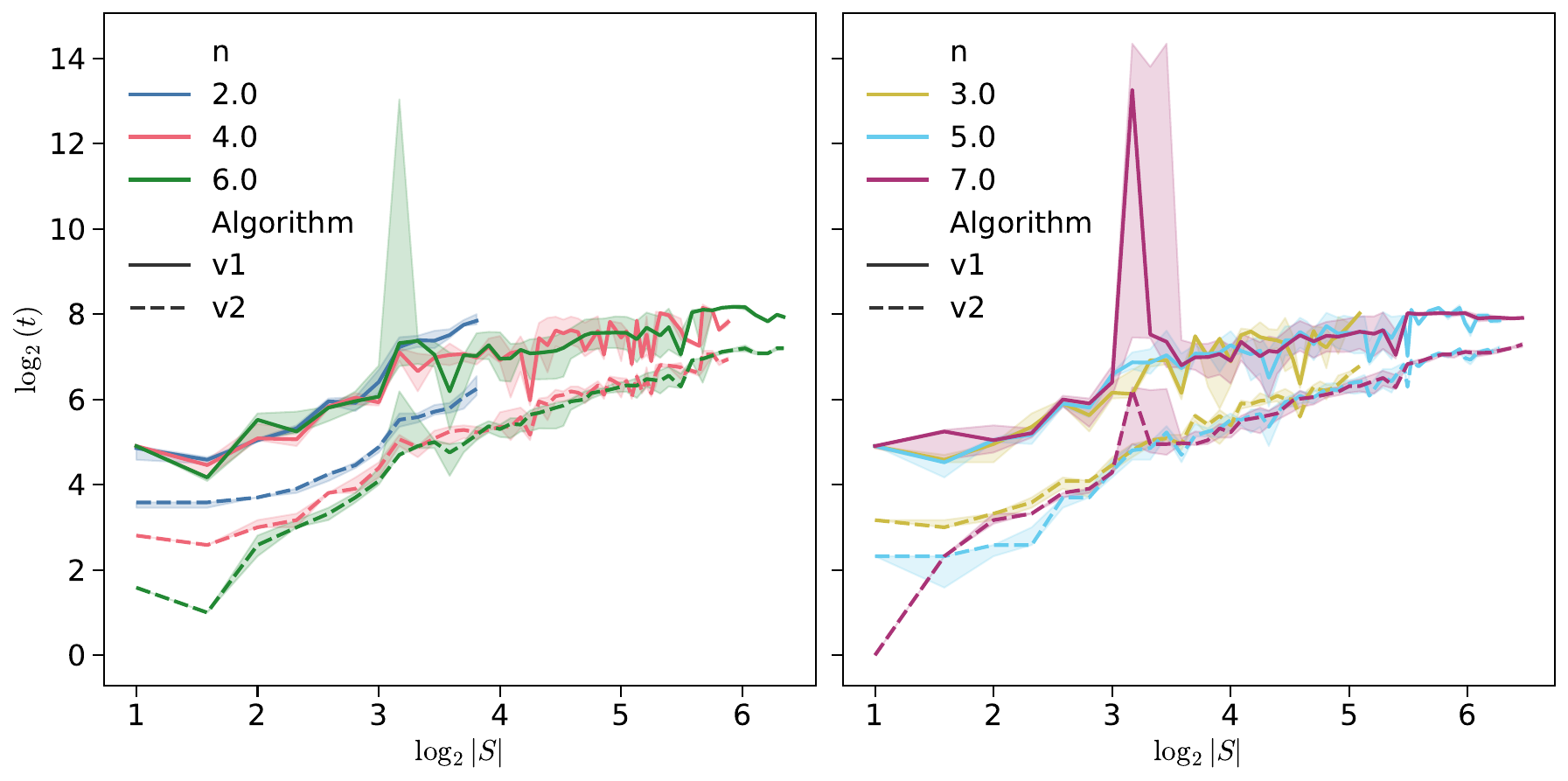}
    \end{subfigure}
    \caption{\emph{Median number of update steps $t$ (log-scaled) for constructing mimicking states for Gibbs test states, shown versus logarithmic support size $\log_2 |S|$}. The results are split into left and right panels by $n$ for visual clarity. Algorithm v2 consistently outperforms v1 across all settings, requiring fewer update steps. This performance gap is especially pronounced where v1 occasionally fails to find feasible solutions (e.g. near $\log_2 |S| \approx 3$ for larger $n$), while v2 terminates early due to its adaptive stopping rule, thereby saving orders of magnitude in computational effort.}
    \label{fig:comp_v1v2_across_alln_gibbs}
\end{figure}

\begin{figure*}[t]
    \centering
    \begin{subfigure}[b]{\textwidth}
        \includegraphics[width=\textwidth]{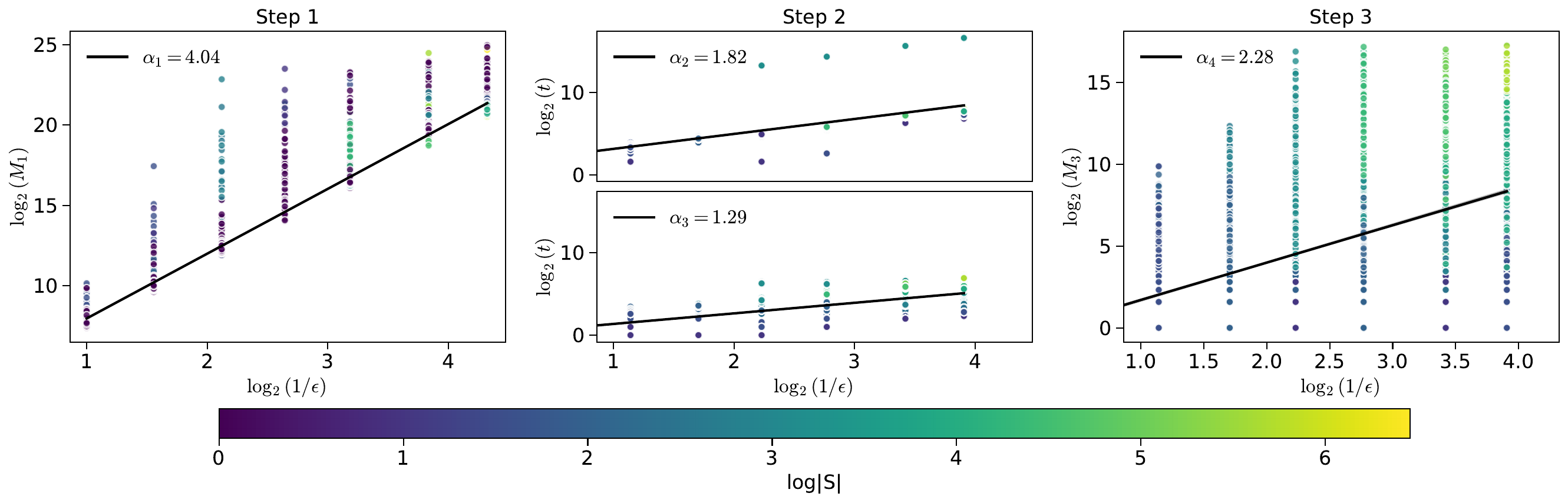}
    \end{subfigure}
    \caption{\emph{Scaling behavior of all algorithm stages for Gibbs states on $n=7$ qubits.} Each column corresponds to a step in the state reconstruction process. In Step 1 (left), the number of samples $M_1$ required for support estimation is plotted versus $\log_{(1/\epsilon)}$, with an overlaid linear fit with the scaling slope of $4.04$. In Step 2 (middle), separate panels for two algorithms display the number of update iterations (log-scaled) against $\log_{(1/\epsilon)}$, with fitted slope of $1.82$ and $1.29$ respectively. Finally, Step 3 (right) shows the sample count $M_3$ required for sign reconstruction versus $\log_2(1/\epsilon)$. The color coding denotes the logarithm of the states support size ($\log_2|S|$). A summary of the scaling behavior for all studied system sizes and states is provided in table~\ref{tab:slope_scaling}.}
    \label{fig:fit_step123_gibbs_n7}
\end{figure*}

\begin{figure*}[t]
    \centering
    \begin{subfigure}[b]{\textwidth}
        \includegraphics[width=\textwidth]{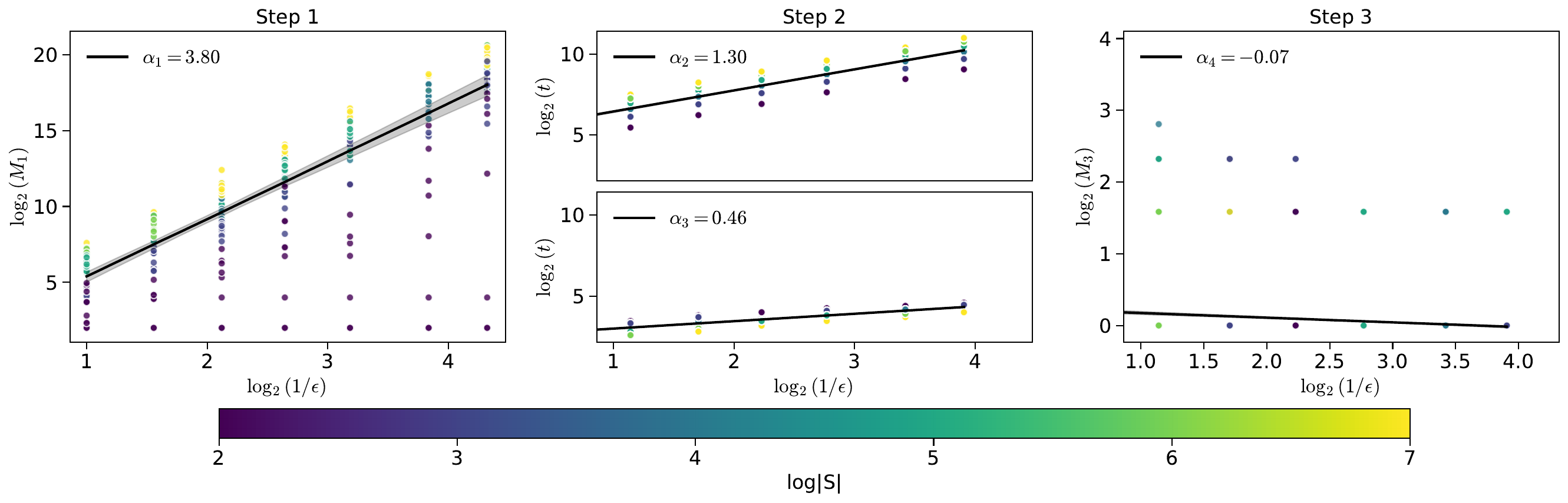}
    \end{subfigure}
    \caption{\emph{Scaling behavior of all algorithm stages for GHZ states across all studied qubit sizes $n=2,\ldots,7$.} Each column corresponds to one of the three stages of the state reconstruction process. In Step 1 (left), the number of samples $M_1$ required for approximate support recovery is plotted versus $\log_2(1/\epsilon)$, with an overlaid linear fit yielding the scaling slope $\alpha_1$. In Step 2 (middle), separate panels for the two algorithm variants show the number of update iterations $t$ (log-scaled) against $\log_2(1/\epsilon)$, along with corresponding fitted slopes $\alpha_2$ and $\alpha_3$. Step 3 (right) shows the number of samples $M_3$ required for accurate sign estimation. The color denotes the logarithm of the support size $\log_2|S|$. Note that the fits shown here were performed over the combined data from all $n$, whereas Table~\ref{tab:slope_scaling} reports scaling exponents fitted separately for each $n$. }
    \label{fig:fit_step123_ghz_alln}
\end{figure*}

In this section, we present and discuss our empirical findings regarding the scaling behavior of each stage of the triply efficient two-copy shadow tomography protocol. We analyze the sample complexity and computational overhead by systematically comparing numerical results to theoretical predictions. The empirical results of our diligent numerical experiments are succinctly summarized in Table~\ref{tab:slope_scaling}. To complement this summary, Fig.~\ref{fig:fit_step123_gibbs_n7} and Fig.~\ref{fig:fit_step123_ghz_alln} provide detailed visualizations of all three reconstruction stages for representative examples: a Gibbs state with $n=7$ qubits and GHZ states across all tested qubit sizes, respectively. These correspond directly to the Gibbs ($n=7$) and GHZ ($n=2,\ldots,7$) rows in Table~\ref{tab:slope_scaling}.
Each row for Gibbs states subsumes 10,000 experiments: for each $n$, we sample 100 random Pauli terms (i.e., 100 Gibbs states), run 10 trials in Stage 1, construct mimicking states using both v1 and v2 (Stage 2), and perform 5 sign estimation trials per construction (Stage 3), yielding $100 \times 10 \times 2 \times 5$ total runs. For GHZ and Zero states, the procedure is identical, but starts from only one input state per $n$, resulting in $1 \times 10 \times 2 \times 5 = 100$ experiments per row. Let us now discuss our findings for each stage and put them into context.

\paragraph*{\textbf{1. Stage -- Magnitude estimation}}
The first stage empirically confirms the expected theoretical sample complexity of $\mathcal{O}(\log|S|/\epsilon^4)$. As shown in Table~\ref{tab:slope_scaling}, numerical simulations demonstrate scaling exponents consistently close to 4 across all considered states and system sizes. This agreement is particularly pronounced for Gibbs and GHZ states. The robust linear relationship between the logarithm of the required samples $\log(M_1)$ and the logarithm of the achieved approximation accuracy $\log(1/\epsilon)$ indicates that successful approximate support recovery via Bell measurements closely aligns with theoretical predictions. To quantify uncertainty, we performed 100 bootstrap resampling trials for each configuration, obtaining distributions over the fitted exponents. The resulting percentiles are reported in Table~\ref{tab:slope_scaling}. In particular, the accuracy scaling of (about) $1/\epsilon^4$ seems to be a necessary and sufficient feature of the Bell sampling approach -- both in theory and in practice.

\paragraph*{\textbf{2. Stage -- Mimicking state construction}}
For the mimicking state construction, we compared the original MMW algorithm (v1) from King et al.\cite{king_triply_2025} with our improved adaptive Hamiltonian Updates algorithm (v2). To quantify scaling with $\epsilon$, we fitted the log-number of update steps ($t$) versus $\log(1/\epsilon)$, where $t$ includes adaptive refinement substeps in v2. Theoretically, the update complexity scales as $\mathcal{O}(n/\epsilon^2)$, ignoring additional sample complexity of $\mathcal{O}(\log(n/\epsilon)/\epsilon^2)$ for the estimation of a Pauli sign in each iteration. Empirically, both algorithms exhibited exponents smaller than the theoretical upper bound ($\alpha_{2,3}<2$), indicating favorable practical scaling.

Table~\ref{tab:slope_scaling} demonstrates the improved scaling behavior of v2 compared to v1 ($\alpha_2 > \alpha_3$). Figure~\ref{fig:comp_v1v2_across_alln_gibbs} further highlights this advantage by showing that v2 consistently requires orders of magnitude fewer updates than v1, particularly around $\log_2(|S|)\approx3$ for larger $n$, where v1 even occasionally fails. A careful analysis traced this back to inaccurate support estimates from Stage 1, heralded by a low Jaccard index close to the threshold 0.9. In such cases, v2's early stopping not only improves robustness but also lays the foundation for adaptive measurement schemes that respond to optimization feedback.
We elaborate this idea further in Sec.~\ref{sec:conclusion} below.

For highly structured stabilizer states (GHZ and computational-basis states), the improvement is even more pronounced, reflecting the algorithm’s ability to effectively leverage structural simplicity.
This is illustrated in Figure~\ref{fig:three_step_process}, panel (f), where the 5-qubit GHZ state requires only a fraction of the update steps compared to an equally sized Gibbs state. At $n=7$, the broader trend is confirmed by Table~\ref{tab:slope_scaling}, where the iteration count exponents drop significantly from around $1.28$ (v1) to approximately $0.39$ (v2). While Gibbs states are more generic, they still show consistent improvement under v2, albeit with a more modest margin. A detailed illustration for the Gibbs state with $n=7$ is shown in Figure~\ref{fig:fit_step123_gibbs_n7} (middle panel).

\paragraph*{\textbf{3. Stage -- Sign estimation}}
With the mimicking state in place, the final stage estimates Pauli signs via Bell-basis measurements on $\rho \otimes \sigma$. This stage again exhibits empirical scaling behavior consistent with theoretical predictions of $\mathcal{O}(\log(|S|)/\epsilon^4)$, particularly for generic Gibbs states.

However, for highly structured GHZ and computational-basis states, the observed numerical exponents are notably lower than theory predictions, as their structured Pauli support simplifies accurate sign recovery even with few measurements. We even observe slightly negative exponents in Table~\ref{tab:slope_scaling}. These are not meaningful and likely result from a few low-$1/\epsilon$ outliers where sign recovery took unusually longer. This behavior is visible in Figure~\ref{fig:fit_step123_ghz_alln} (right panel), where the otherwise flat trend in the log-log fit is distorted by a small number of outliers at low $1/\epsilon$. Fits for the zero state are visually nearly identical to those of the GHZ state and are therefore omitted.
It is also worthwhile to point out that sign estimation generally required significantly fewer measurements than the support identification in stage 1. This difference is also clearly visible in Fig.\ref{fig:fit_step123_gibbs_n7}, where the number of samples required for Stage~1 exceeds that of Stage~3 by several orders of magnitude.
This indicates that once the support is accurately identified and the mimicking state is constructed, the subsequent sign determination is relatively easy by comparison.

Overall, our empirical evaluations corroborate theoretical scaling predictions across all three stages of triply-efficient Pauli shadow tomography, while additionally revealing notable practical improvements achieved by our adaptive Hamiltonian Updates strategy (v2). This enhanced subroutine, in particular, represents a meaningful step forward in practical shadow tomography implementations, effectively addressing computational bottlenecks and improving convergence stability and efficiency.

\section{Conclusion and Outlook}\label{sec:conclusion}

In this work, we conducted the first empirical evaluation of
a quantum-enhanced Pauli shadow tomography protocol that is triply efficient~\cite{king_triply_2025} in the sense that (i) it requires small sample complexity, (ii) it requires tractable joint measurements on few state copies and (iii) the classical co-processing overhead can be benign. This protocol operates in three stages and our results confirm that the empirical sample complexity aligns closely with theoretical predictions for generic states, exemplified by random Gibbs states, and notably exhibit slightly improved scaling behavior for highly structured states, exemplified by stabilizer states. 

Moreover, by leveraging recent insights from advanced quantum-inspired convex optimization techniques developed by us and collaborators in Cologne and Braunschweig~\cite{henze_solving_2025}, we significantly improved the second stage of the original protocol. This stage hinges on a matrix multiplicative weight (MMW) subroutine and our improved variant -- called Hamiltonian Updates -- demonstrates substantial practical advantages, such as reduced iteration counts and more robust numerical convergence. Hence, our findings highlight promising avenues for further optimizing hybrid quantum-classical approaches for Pauli shadow tomography -- a well-motivated quantum learning task with many applications ranging from many-body physics to quantum chemistry.

Additionally, our enhanced MMW algorithm can serve as a practical, data-driven feedback mechanism for adaptively determining the necessity of additional samples due to its faster signaling of non-convergence, further optimizing the efficiency of the shadow tomography protocol.
Our findings underscore that triply efficient shadow tomography protocols exhibit favorable constants and performance in practice, suggesting significant potential for efficient experimental implementation.

Looking forward, several intriguing directions present themselves. Further enhancements of the MMW update strategy, such as integrating momentum-based and batching approaches, could yield additional computational efficiency. Exploring extensions beyond Pauli observables towards broader classes of measurements and states also represents a promising avenue, potentially expanding the applicability of shadow tomography techniques to an even wider range of quantum learning tasks.

\section*{Acknowledgments}

The authors acknowledge financial support from the Austrian Science Fund via the SFB BeyondC (Grant No. F7107-N38), the Austrian Research Promotion Agency (FFG) via the Quantum Austria project QuantumReady (FFG 896217), and the European Research Council (ERC Grant Agreement No.
101117138, q-shadows).
The authors also thank Johannes Brandstetter, Sepp Hochreiter and G\"unter Klambauer from the Institute for Machine Learning at Johannes Kepler University Linz for granting access to their computational resources, which were essential for carrying out this work.

\bibliographystyle{IEEEtran}
\bibliography{references}

\begin{thebibliography}{10}
\providecommand{\url}[1]{#1}
\csname url@samestyle\endcsname
\providecommand{\newblock}{\relax}
\providecommand{\bibinfo}[2]{#2}
\providecommand{\BIBentrySTDinterwordspacing}{\spaceskip=0pt\relax}
\providecommand{\BIBentryALTinterwordstretchfactor}{4}
\providecommand{\BIBentryALTinterwordspacing}{\spaceskip=\fontdimen2\font plus
\BIBentryALTinterwordstretchfactor\fontdimen3\font minus \fontdimen4\font\relax}
\providecommand{\BIBforeignlanguage}[2]{{%
\expandafter\ifx\csname l@#1\endcsname\relax
\typeout{** WARNING: IEEEtran.bst: No hyphenation pattern has been}%
\typeout{** loaded for the language `#1'. Using the pattern for}%
\typeout{** the default language instead.}%
\else
\language=\csname l@#1\endcsname
\fi
#2}}
\providecommand{\BIBdecl}{\relax}
\BIBdecl

\bibitem{HKP21}
\BIBentryALTinterwordspacing
H.-Y. Huang, R.~Kueng, and J.~Preskill, ``Information-{Theoretic} {Bounds} on {Quantum} {Advantage} in {Machine} {Learning},'' \emph{Physical Review Letters}, vol. 126, no.~19, p. 190505, May 2021, publisher: American Physical Society. [Online]. Available: \url{https://link.aps.org/doi/10.1103/PhysRevLett.126.190505}
\BIBentrySTDinterwordspacing

\bibitem{king_triply_2025}
\BIBentryALTinterwordspacing
R.~King, D.~Gosset, R.~Kothari, and R.~Babbush, ``Triply efficient shadow tomography,'' in \emph{Proceedings of the 2025 Annual {ACM}-{SIAM} Symposium on Discrete Algorithms ({SODA})}, ser. Proceedings.\hskip 1em plus 0.5em minus 0.4em\relax Society for Industrial and Applied Mathematics, pp. 914--946. [Online]. Available: \url{https://epubs.siam.org/doi/abs/10.1137/1.9781611978322.27}
\BIBentrySTDinterwordspacing

\bibitem{henze_solving_2025}
\BIBentryALTinterwordspacing
F.~Henze, V.~Tran, B.~Ostermann, R.~Kueng, T.~d. Wolff, and D.~Gross, ``Solving quadratic binary optimization problems using quantum {SDP} methods: Non-asymptotic running time analysis.'' [Online]. Available: \url{http://arxiv.org/abs/2502.15426}
\BIBentrySTDinterwordspacing

\bibitem{BCG13}
\BIBentryALTinterwordspacing
K.~Banaszek, M.~Cramer, and D.~Gross, ``Focus on quantum tomography,'' \emph{New Journal of Physics}, vol.~15, no.~12, p. 125020, 12 2013. [Online]. Available: \url{https://dx.doi.org/10.1088/1367-2630/15/12/125020}
\BIBentrySTDinterwordspacing

\bibitem{KRT17}
\BIBentryALTinterwordspacing
R.~Kueng, H.~Rauhut, and U.~Terstiege, ``Low rank matrix recovery from rank one measurements,'' \emph{Applied and Computational Harmonic Analysis}, vol.~42, no.~1, pp. 88--116, 2017. [Online]. Available: \url{https://www.sciencedirect.com/science/article/pii/S1063520315001037}
\BIBentrySTDinterwordspacing

\bibitem{DW16}
\BIBentryALTinterwordspacing
R.~O'Donnell and J.~Wright, ``Efficient quantum tomography,'' in \emph{Proceedings of the Forty-Eighth Annual ACM Symposium on Theory of Computing}, ser. STOC '16.\hskip 1em plus 0.5em minus 0.4em\relax New York, NY, USA: Association for Computing Machinery, 2016, p. 899–912. [Online]. Available: \url{https://doi.org/10.1145/2897518.2897544}
\BIBentrySTDinterwordspacing

\bibitem{HHJWN16}
\BIBentryALTinterwordspacing
J.~Haah, A.~W. Harrow, Z.~Ji, X.~Wu, and N.~Yu, ``Sample-optimal tomography of quantum states,'' in \emph{Proceedings of the Forty-Eighth Annual ACM Symposium on Theory of Computing}, ser. STOC '16.\hskip 1em plus 0.5em minus 0.4em\relax New York, NY, USA: Association for Computing Machinery, 2016, p. 913–925. [Online]. Available: \url{https://doi.org/10.1145/2897518.2897585}
\BIBentrySTDinterwordspacing

\bibitem{FBK21}
\BIBentryALTinterwordspacing
D.~S. Fran{\c{c}}a, F.~G. S.~L. Brand{\~{a}}o, and R.~Kueng, ``Fast and robust quantum state tomography from few basis measurements,'' in \emph{16th Conference on the Theory of Quantum Computation, Communication and Cryptography, {TQC} 2021, July 5-8, 2021, Virtual Conference}, ser. LIPIcs, M.~Hsieh, Ed., vol. 197.\hskip 1em plus 0.5em minus 0.4em\relax Schloss Dagstuhl - Leibniz-Zentrum f{\"{u}}r Informatik, 2021, pp. 7:1--7:13. [Online]. Available: \url{https://doi.org/10.4230/LIPIcs.TQC.2021.7}
\BIBentrySTDinterwordspacing

\bibitem{AL25}
\BIBentryALTinterwordspacing
A.~Nayak and A.~Lowe, ``Lower bounds for learning quantum states with single-copy measurements,'' \emph{ACM Transactions on Computation Theory}, vol.~17, no.~1, Mar. 2025. [Online]. Available: \url{https://doi.org/10.1145/3717450}
\BIBentrySTDinterwordspacing

\bibitem{Aar18}
\BIBentryALTinterwordspacing
S.~Aaronson, ``Shadow tomography of quantum states,'' in \emph{Proceedings of the 50th Annual {ACM} {SIGACT} Symposium on Theory of Computing, {STOC} 2018, Los Angeles, CA, USA, June 25-29, 2018}, I.~Diakonikolas, D.~Kempe, and M.~Henzinger, Eds.\hskip 1em plus 0.5em minus 0.4em\relax {ACM}, 2018, pp. 325--338. [Online]. Available: \url{https://doi.org/10.1145/3188745.3188802}
\BIBentrySTDinterwordspacing

\bibitem{BD21}
\BIBentryALTinterwordspacing
C.~Badescu and R.~O'Donnell, ``Improved quantum data analysis,'' in \emph{{STOC} '21: 53rd Annual {ACM} {SIGACT} Symposium on Theory of Computing, Virtual Event, Italy, June 21-25, 2021}, S.~Khuller and V.~V. Williams, Eds.\hskip 1em plus 0.5em minus 0.4em\relax {ACM}, 2021, pp. 1398--1411. [Online]. Available: \url{https://doi.org/10.1145/3406325.3451109}
\BIBentrySTDinterwordspacing

\bibitem{CPF+10}
\BIBentryALTinterwordspacing
M.~Cramer, M.~B. Plenio, S.~T. Flammia, R.~Somma, D.~Gross, S.~D. Bartlett, O.~Landon-Cardinal, D.~Poulin, and Y.-K. Liu, ``Efficient quantum state tomography,'' vol.~1, no.~1, p. 149, publisher: Nature Publishing Group. [Online]. Available: \url{https://www.nature.com/articles/ncomms1147}
\BIBentrySTDinterwordspacing

\bibitem{TMC+18}
\BIBentryALTinterwordspacing
G.~Torlai, G.~Mazzola, J.~Carrasquilla, M.~Troyer, R.~Melko, and G.~Carleo, ``Neural-network quantum state tomography,'' vol.~14, no.~5, pp. 447--450, publisher: Nature Publishing Group. [Online]. Available: \url{https://www.nature.com/articles/s41567-018-0048-5}
\BIBentrySTDinterwordspacing

\bibitem{HKT22}
\BIBentryALTinterwordspacing
J.~Haah, R.~Kothari, and E.~Tang, ``Optimal learning of quantum hamiltonians from high-temperature gibbs states,'' in \emph{63rd {IEEE} Annual Symposium on Foundations of Computer Science, {FOCS} 2022, Denver, CO, USA, October 31 - November 3, 2022}.\hskip 1em plus 0.5em minus 0.4em\relax {IEEE}, 2022, pp. 135--146. [Online]. Available: \url{https://doi.org/10.1109/FOCS54457.2022.00020}
\BIBentrySTDinterwordspacing

\bibitem{RFOW24}
\BIBentryALTinterwordspacing
C.~Rouzé, D.~Stilck~França, E.~Onorati, and J.~D. Watson, ``Efficient learning of ground and thermal states within phases of matter,'' vol.~15, no.~1, p. 7755, publisher: Nature Publishing Group. [Online]. Available: \url{https://www.nature.com/articles/s41467-024-51439-x}
\BIBentrySTDinterwordspacing

\bibitem{MD19}
\BIBentryALTinterwordspacing
J.~Morris and B.~Dakić, ``Selective quantum state tomography.'' [Online]. Available: \url{http://arxiv.org/abs/1909.05880}
\BIBentrySTDinterwordspacing

\bibitem{PK19}
\BIBentryALTinterwordspacing
M.~Paini and A.~Kalev, ``An approximate description of quantum states.'' [Online]. Available: \url{http://arxiv.org/abs/1910.10543}
\BIBentrySTDinterwordspacing

\bibitem{HKP20}
\BIBentryALTinterwordspacing
H.-Y. Huang, R.~Kueng, and J.~Preskill, ``Predicting many properties of a quantum system from very few measurements,'' vol.~16, no.~10, pp. 1050--1057, publisher: Nature Publishing Group. [Online]. Available: \url{https://www.nature.com/articles/s41567-020-0932-7}
\BIBentrySTDinterwordspacing

\bibitem{VTI20}
\BIBentryALTinterwordspacing
V.~Verteletskyi, T.-C. Yen, and A.~F. Izmaylov, ``Measurement optimization in the variational quantum eigensolver using a minimum clique cover,'' \emph{The Journal of Chemical Physics}, vol. 152, no.~12, p. 124114, 03 2020. [Online]. Available: \url{https://doi.org/10.1063/1.5141458}
\BIBentrySTDinterwordspacing

\bibitem{HKP21b}
\BIBentryALTinterwordspacing
H.-Y. Huang, R.~Kueng, and J.~Preskill, ``Efficient estimation of pauli observables by derandomization,'' \emph{Physical Review Letters}, vol. 127, p. 030503, 7 2021. [Online]. Available: \url{https://link.aps.org/doi/10.1103/PhysRevLett.127.030503}
\BIBentrySTDinterwordspacing

\bibitem{GK25}
\BIBentryALTinterwordspacing
A.~Gresch and M.~Kliesch, ``Guaranteed efficient energy estimation of quantum many-body hamiltonians using {ShadowGrouping},'' vol.~16, no.~1, p. 689, publisher: Nature Publishing Group. [Online]. Available: \url{https://www.nature.com/articles/s41467-024-54859-x}
\BIBentrySTDinterwordspacing

\bibitem{KKK+24}
\BIBentryALTinterwordspacing
K.~V. Kirk, C.~Kokail, J.~Kunjummen, H.-Y. Hu, Y.~Teng, M.~Cain, J.~Taylor, S.~F. Yelin, H.~Pichler, and M.~Lukin, ``Derandomized shallow shadows: Efficient pauli learning with bounded-depth circuits.'' [Online]. Available: \url{http://arxiv.org/abs/2412.18973}
\BIBentrySTDinterwordspacing

\bibitem{CCHL21}
\BIBentryALTinterwordspacing
S.~Chen, J.~Cotler, H.~Huang, and J.~Li, ``Exponential separations between learning with and without quantum memory,'' in \emph{62nd {IEEE} Annual Symposium on Foundations of Computer Science, {FOCS} 2021, Denver, CO, USA, February 7-10, 2022}.\hskip 1em plus 0.5em minus 0.4em\relax {IEEE}, 2021, pp. 574--585. [Online]. Available: \url{https://doi.org/10.1109/FOCS52979.2021.00063}
\BIBentrySTDinterwordspacing

\bibitem{HBC+22}
\BIBentryALTinterwordspacing
H.-Y. Huang, M.~Broughton, J.~Cotler, S.~Chen, J.~Li, M.~Mohseni, H.~Neven, R.~Babbush, R.~Kueng, J.~Preskill, and J.~R. {McClean}, ``Quantum advantage in learning from experiments,'' vol. 376, no. 6598, pp. 1182--1186, publisher: American Association for the Advancement of Science. [Online]. Available: \url{https://www.science.org/doi/full/10.1126/science.abn7293}
\BIBentrySTDinterwordspacing

\bibitem{CGY24}
\BIBentryALTinterwordspacing
S.~Chen, W.~Gong, and Q.~Ye, ``Optimal tradeoffs for estimating pauli observables,'' in \emph{65th {IEEE} Annual Symposium on Foundations of Computer Science, {FOCS} 2024, Chicago, IL, USA, October 27-30, 2024}.\hskip 1em plus 0.5em minus 0.4em\relax {IEEE}, 2024, pp. 1086--1105. [Online]. Available: \url{https://doi.org/10.1109/FOCS61266.2024.00072}
\BIBentrySTDinterwordspacing

\bibitem{SK14}
\BIBentryALTinterwordspacing
R.~Koenig and J.~A. Smolin, ``How to efficiently select an arbitrary clifford group element,'' \emph{Journal of Mathematical Physics}, vol.~55, no.~12, p. 122202, 12 2014. [Online]. Available: \url{https://doi.org/10.1063/1.4903507}
\BIBentrySTDinterwordspacing

\bibitem{CBTW17}
\BIBentryALTinterwordspacing
P.~J. Coles, M.~Berta, M.~Tomamichel, and S.~Wehner, ``Entropic uncertainty relations and their applications,'' \emph{Review of Modern Physics}, vol.~89, p. 015002, 2 2017. [Online]. Available: \url{https://link.aps.org/doi/10.1103/RevModPhys.89.015002}
\BIBentrySTDinterwordspacing

\bibitem{TOVPV11}
\BIBentryALTinterwordspacing
K.~Temme, T.~J. Osborne, K.~G. Vollbrecht, D.~Poulin, and F.~Verstraete, ``Quantum metropolis sampling,'' vol. 471, no. 7336, pp. 87--90, publisher: Nature Publishing Group. [Online]. Available: \url{https://www.nature.com/articles/nature09770}
\BIBentrySTDinterwordspacing

\bibitem{RFA24}
\BIBentryALTinterwordspacing
C.~Rouzé, D.~S. França, and {\'A}.~M. Alhambra, ``Optimal quantum algorithm for gibbs state preparation.'' [Online]. Available: \url{http://arxiv.org/abs/2411.04885}
\BIBentrySTDinterwordspacing

\bibitem{arora_multiplicative_2012}
\BIBentryALTinterwordspacing
S.~Arora, E.~Hazan, and S.~Kale, ``The multiplicative weights update method: a meta-algorithm and applications,'' vol.~8, pp. 121--164, number: 6 Publisher: Theory of Computing. [Online]. Available: \url{https://theoryofcomputing.org/articles/v008a006/}
\BIBentrySTDinterwordspacing

\bibitem{brandao_faster_2022}
\BIBentryALTinterwordspacing
F.~G. S.~L. Brandão, R.~Kueng, and D.~S. França, ``Faster quantum and classical {SDP} approximations for quadratic binary optimization,'' vol.~6, p. 625, publisher: Verein zur Förderung des Open Access Publizierens in den Quantenwissenschaften. [Online]. Available: \url{https://quantum-journal.org/papers/q-2022-01-20-625/}
\BIBentrySTDinterwordspacing

\bibitem{greenberger_going_1989}
\BIBentryALTinterwordspacing
D.~M. Greenberger, M.~A. Horne, and A.~Zeilinger, ``Going beyond bell’s theorem,'' in \emph{Bell’s Theorem, Quantum Theory and Conceptions of the Universe}, M.~Kafatos, Ed.\hskip 1em plus 0.5em minus 0.4em\relax Springer Netherlands, pp. 69--72. [Online]. Available: \url{https://doi.org/10.1007/978-94-017-0849-4_10}
\BIBentrySTDinterwordspacing

\bibitem{harris2020array}
\BIBentryALTinterwordspacing
C.~R. Harris, K.~J. Millman, S.~J. van~der Walt, R.~Gommers, P.~Virtanen, D.~Cournapeau, E.~Wieser, J.~Taylor, S.~Berg, N.~J. Smith, R.~Kern, M.~Picus, S.~Hoyer, M.~H. van Kerkwijk, M.~Brett, A.~Haldane, J.~F. del R{\'{i}}o, M.~Wiebe, P.~Peterson, P.~G{\'{e}}rard-Marchant, K.~Sheppard, T.~Reddy, W.~Weckesser, H.~Abbasi, C.~Gohlke, and T.~E. Oliphant, ``Array programming with {NumPy},'' \emph{Nature}, vol. 585, no. 7825, pp. 357--362, Sep. 2020. [Online]. Available: \url{https://doi.org/10.1038/s41586-020-2649-2}
\BIBentrySTDinterwordspacing

\bibitem{2020SciPy-NMeth}
P.~Virtanen, R.~Gommers, T.~E. Oliphant, M.~Haberland, T.~Reddy, D.~Cournapeau, E.~Burovski, P.~Peterson, W.~Weckesser, J.~Bright, S.~J. {van der Walt}, M.~Brett, J.~Wilson, K.~J. Millman, N.~Mayorov, A.~R.~J. Nelson, E.~Jones, R.~Kern, E.~Larson, C.~J. Carey, {\.I}.~Polat, Y.~Feng, E.~W. Moore, J.~{VanderPlas}, D.~Laxalde, J.~Perktold, R.~Cimrman, I.~Henriksen, E.~A. Quintero, C.~R. Harris, A.~M. Archibald, A.~H. Ribeiro, F.~Pedregosa, P.~{van Mulbregt}, and {SciPy 1.0 Contributors}, ``{{SciPy} 1.0: Fundamental Algorithms for Scientific Computing in Python},'' \emph{Nature Methods}, vol.~17, pp. 261--272, 2020.

\bibitem{lam2015numba}
\BIBentryALTinterwordspacing
S.~K. Lam, A.~Pitrou, and S.~Seibert, ``Numba: a {LLVM}-based python {JIT} compiler,'' in \emph{Proceedings of the Second Workshop on the {LLVM} Compiler Infrastructure in {HPC}}, ser. {LLVM} '15.\hskip 1em plus 0.5em minus 0.4em\relax Association for Computing Machinery, pp. 1--6. [Online]. Available: \url{https://dl.acm.org/doi/10.1145/2833157.2833162}
\BIBentrySTDinterwordspacing

\bibitem{cupy_learningsys2017}
\BIBentryALTinterwordspacing
R.~Okuta, Y.~Unno, D.~Nishino, S.~Hido, and C.~Loomis, ``Cupy: A numpy-compatible library for nvidia gpu calculations,'' in \emph{Proceedings of Workshop on Machine Learning Systems (LearningSys) in The Thirty-first Annual Conference on Neural Information Processing Systems (NIPS)}, 2017. [Online]. Available: \url{http://learningsys.org/nips17/assets/papers/paper_16.pdf}
\BIBentrySTDinterwordspacing

\bibitem{tran_one_2025}
V.~T. Tran and R.~Kueng, ``{Code for "One, Two, Three: One Empirical Evaluation of a Two-Copy Shadow Tomography Scheme with Triple Efficiency"},'' Online. Available: \url{https://github.com/VietTralala/123-triply-eff}, 2025.

\end{thebibliography}

\end{document}